\documentclass{article}
\newcommand{\FTS}{FeTe$_x$Se$_{1-x}$}
\newcommand{\alum}{Al$_2$O$_3$}
\usepackage{hyperref}
\usepackage[super,comma,compress]{natbib}
\usepackage{graphicx}
\usepackage[a4paper, total={6.5in, 10.5in}]{geometry}
\usepackage{float}
\usepackage{hyperref}
\hypersetup{
    colorlinks=true,
    linkcolor=blue,
citecolor=black}

\usepackage{setspace}
\begin{document}

\title{Chip-Integrated Vortex Manipulation}

\date{}
\maketitle

\author{I.Keren$^1$,
A.Gutfreund$^1$,
A.Noah$^1$,
N.Friedman$^1$
A.Di Bernardo$^2$,
H.Steinberg$^{1,3}$,
Y.Anahory$^{1,3}$
\smallskip 
\\
$^1$Racah Institute of Physics, The Hebrew University, Jerusalem 91904, Israel
\\
$^2$Department of Physics, University of Konstanz, Universitätstrasse 10, 78457 Konstanz
\\
$^3$Racah Institute of Physics and the center for NanoScience and Nanotechnology, The Hebrew University, Jerusalem 91904, Israel}

\bigskip

Corresponding Authors: yonathan.anahory@mail.huji.ac.il, hadar@phys.huji.ac.il

\smallskip

\textbf{Abrikosov vortices have long been considered as means to encode classical information in low-temperature logic circuits~\cite{Reichhardt2004} and memory devices.~\cite{Hebard1978,Bachtold1979,Parisi1982} Although it is possible to control individual vortices using local probes~\cite{Moler2002,Straver2008,Auslaender2009,Kalisky2011,Ge2016,Veshchunov2016,Beena2016}, scalability towards the control of multiple vortices remains challenging. 
Vortex logic devices require means to shuttle selected vortices reliably over long distances between engineered pinning potentials. Concomitantly, all other vortices should remain fixed to their precise locations. Here we demonstrate such capabilities using Nb loops patterned below a NbSe$_2$ layer. SQUID-on-Tip (SOT) microscopy reveals that the loops can position vortices in sites designated to a precision better than 100~nm; they can realize “push” and “pull” operations of vortices as far as  \hbox{3 $\mathrm{\mu}$m}. Successive application of such operations shuttles a vortex
between adjacent loops. Our results may be used as means to integrate vortices in future quantum circuitry. Strikingly, we are able to demonstrate a winding operation. Such winding, if realized in topological superconductors, is considered an essential part of future topological quantum information processing.~\cite{Read2000, Ivanov2001, Ady2004,Kitaev2006, Fu_Kane, Nayak_RMP_2008}
}

\smallskip
Scalable single-vortex control has long been missing in the nanoscale superconducting device toolbox. Such control could be used in a number of applications. The presence or the absence of an Abrikosov vortex within a Josephson device was suggested as a classical computation element~\cite{Golod2015}. In contrast, in quantum coherent circuitry, vortex removal from undesired positions could overcome vortex induced dissipation and decoherence.~\cite{Plourde2009,Song2009,Plourde2014} Vortex manipulation is achievable by locally reducing the gap $\Delta$, using current injection~\cite{Ge2016,Kalcheim2017}, heat~\cite{Ge2016,Veshchunov2016} or mechanical pressure~\cite{Beena2016}. However, trapping vortices by reducing the gap is limited to  a short range. It can only attract vortices (and not repel them), and it drives the system out of equilibrium. Another avenue, is to exert a long-range Lorentz force which can either attract or repel vortices.~\cite{VanDeVondel2006,Straver2008,Kalisky2009,Auslaender2009,Zeldov2015}

In this work we develop a dynamic vortex control method. It is based on loop elements which may be placed above or below the target superconductor. A broad loop with a finely defined inner diameter combines the high current required to exert magnetic forces over long ranges, with a finely-resolved pinning potential at the center. Use of a loop material with a low penetration length $\lambda$, locally expels vortices and creates an effective exclusion zone for vortices within the target superconductor. The loops are designed with a directional asymmetry, exerting a directional force along a pre-determined vector.

\begin{figure}[H]
\centering

\includegraphics[width=\textwidth]{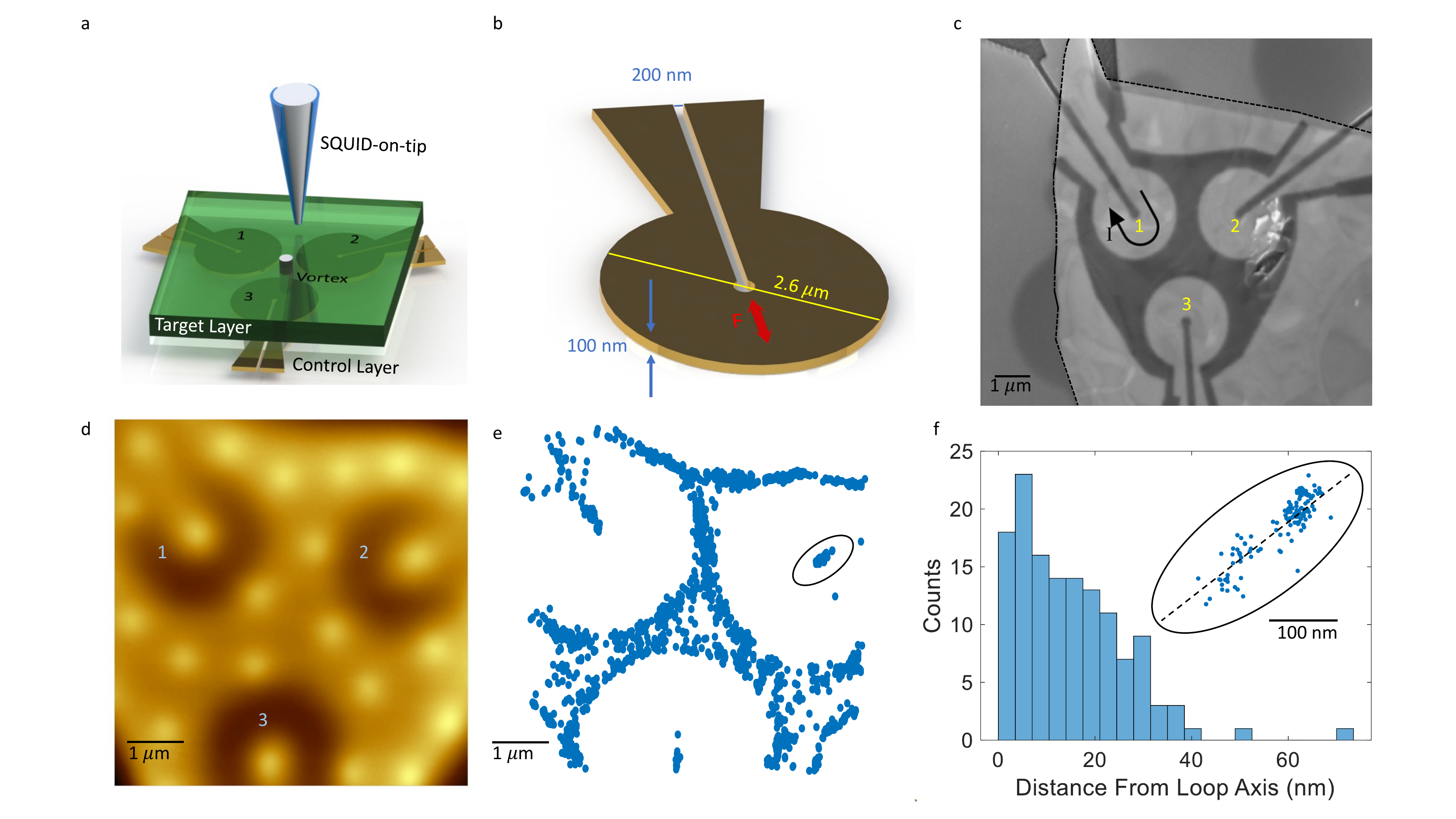}

\caption{\textbf{Device Parameters - } \textbf{a} Schematic drawing of the device. The Nb loops (gold) that are etched in the control layer are separated from a $\sim$ 50 nm thick NbSe$_2$ target layer (green) by a $\sim$ 15 nm thick \alum\ layer (transparent). A vortex is illustrated within the target layer (white cylinder). The SQUID-on-tip (SOT) probe is represented above the device as it images the device. \textbf{b} Loop geometry. The inner diameter is 200 nm (blue line) and the thickness is 100 nm. The outer diameter is 2.6 $\mu$m (yellow line). The direction of the total force (F) applied by the loop on the vortex is marked by the red arrow and is parallel to the slit.  \textbf{c} A top view SEM image of the device. The loops are numbered for future reference. The NbSe$_2$ flake covers all three loops and is highlighted by the black line. The NbSe$_2$ is thin enough to be transparent to the electron beam. The curved arrow indicates the current direction running through Loop 1. \textbf{d} A field cooled state with high vortex density. All three loop centers are occupied, each by a single vortex, while no vortices reside overlapping the loop conductor. \textbf{e} A compilation of 2370 observed vortices in 500 images. The dots within the black ellipse are used to create the histogram in \textbf{f}. \textbf{f} A histogram of the points within the black ellipse in e. This histogram counts the distances of vortices captured within Loop 2, from the loop axis (dotted black line in the inset where a zoomed in scatter is diplayed) resulting in standard deviation of $\sim$ 25 nm. }

\label{fig:Device}
\end{figure}

 Fig.~\ref{fig:Device}a presents the basic paradigm of the experiment, where we use a high resolution SOT microscope~\cite{Zeldov2013,Anahory2020} to magnetically image the device, where the full vortex-control functionality is realized. The device consists of a bottom layer of 100 nm thick Nb superconducting loops fabricated using standard reactive ion etching (see methods and Supplementary Note 1). A single loop, shown schematically in Fig.~\ref{fig:Device}b, has an inner diameter of 200 nm and outer diameter of 2.6 $\mu m$. An SEM image of the loops is shown in Fig.~\ref{fig:Device}c. We refer to these loops as the \textit{control layer}. An Al$_{2}$O$_{3}$ layer ($\sim$15 nm) is deposited on the control layer to electrically decouple the loops from the top superconducting layer where the vortices, which we want to manipulate, reside. Here we use an NbSe$_{2}$ flake positioned using the standard polycarbonate (PC) flake transfer method. We refer to the latter as the \textit{target layer}. The target layer is depicted schematically as a green square in Fig.~\ref{fig:Device}a and imaged in Fig.~\ref{fig:Device}c where the flake edges are outlined with a dashed black line. The loops are numbered for later reference (see Fig.~\ref{fig:Device}a,c and d). Detailed design considerations are discussed in Supplementary Note 2.

Upon application of a current $I$ through any one of the loops, a local magnetic field is generated around its center. This field interacts with the vortex via the Lorentz force $\textbf{F}=\int(\textbf{j} \times \textbf{B}) d^{3}x$. Here $\mathbf{B}$ is the magnetic field applied by the loop and $\mathbf{j}$ is the current density around the vortex core (see Supplementary Note 3).  
Importantly, the direction of the net force applied upon the vortex is sketched as a red arrow in Fig.~\ref{fig:Device}b. We orient all the loops with their force vector pointing towards the central region between them.

In Fig.~\ref{fig:Device}d, we present a SOT magnetic image of the device following a local field cool at 1.4 mT. To do this, the field is applied while imposing $I > I_c$ in any of the loops ($I_c$ being the critical current of the loop). This way, the dissipative current heats the target layer above its critical temperature allowing vortex entry as the loop current is lowered below $I_c$, allowing the target layer to cool. The image which exhibits a scatter of multiple vortices reveals two important features. First, the image shows a vortex at the center of each of the three loops. This is reproducible throughout our measurements, including images taken at higher vortex densities. The reason vortices tend to favor the loop center is related to the second observation seen in the figure - namely the existence of an exclusion zone exactly where the NbSe$_2$ target layer is placed above the Nb loop. In those regions, a weaker magnetic signal is detected, and no vortices are found. The exclusion zone is created since threading a flux line through the loops requires penetrating a second superconductor.

We emphasise these observations by compiling vortex positions extracted from 500 images (Fig.~\ref{fig:Device}e) where each of the 2370 detected vortices is represented by a point. This compilation includes images at high vortex density, low vortex density and images which are taken following the application of force on the vortices using the loops. Fig.~\ref{fig:Device}e confirms that vortices shun the exclusion zone. The outlying cases where vortices are found within the exclusion zone represent $\sim$ 0.1 percent of all detected vortices. 

Our design confines the vortices to a very small and elongated region aligned with the loop axis and located near the loop center.
A zoomed-in image of the vortex scatter appears in the inset to Figure ~\ref{fig:Device}f, where the main panel depicts a histogram of the distances from Loop 2 axis for all the vortices detected within a 100 nm radius of its center.
We find the vortices scatter within a standard deviation of $\sim 25$ nm from the axis. 
Such reliable control over vortex positioning, confirmed by the uniquely high spatial resolution of the SOT, is important for future vortex-control protocols.

\begin{figure}[H]
\center

\includegraphics[width=\linewidth]{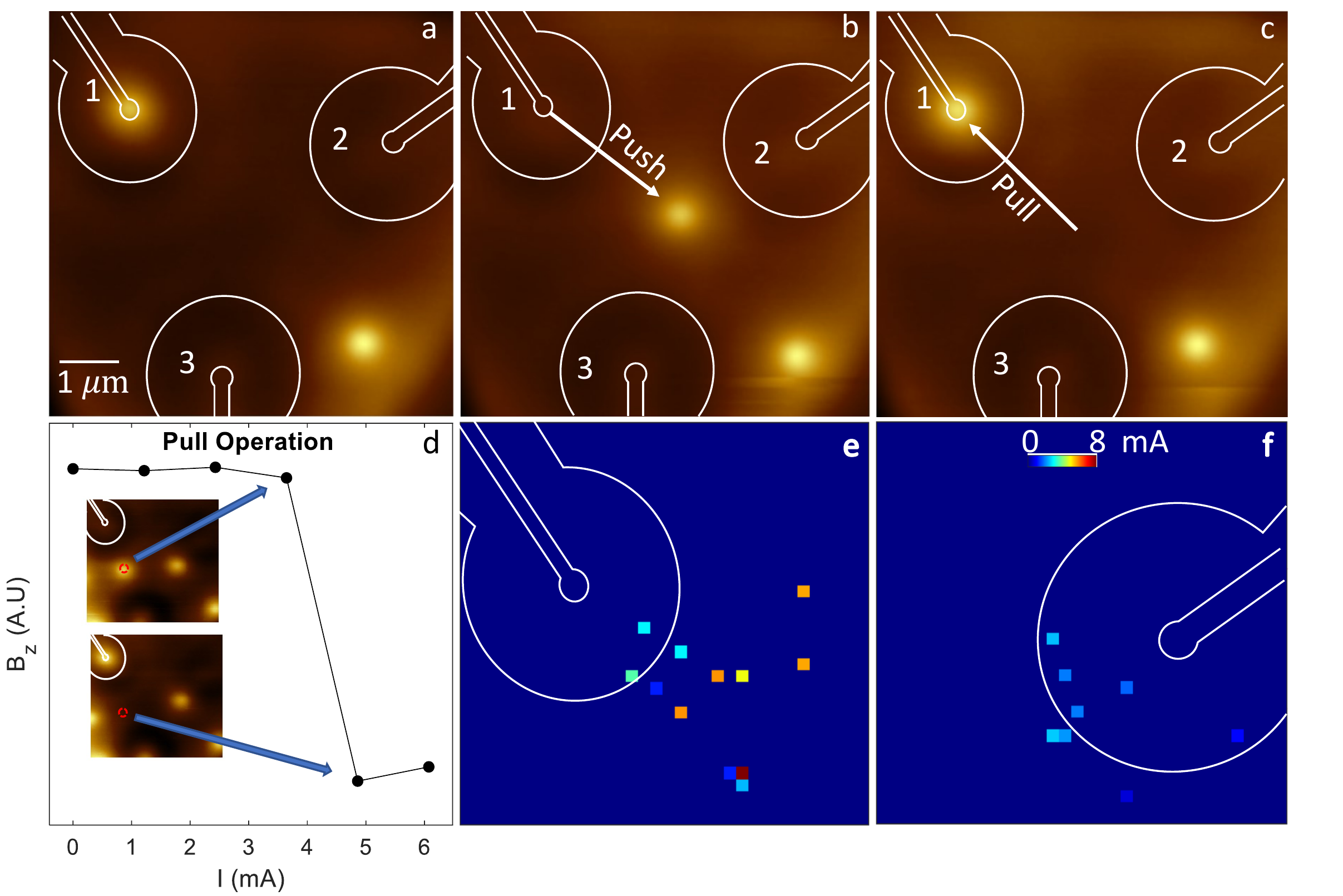}
\caption{\textbf{Basic vortex manipulation - } \textbf{a} The initial state of the system - Loop 1 is occupied. \textbf{b} After passing 2.2 mA through Loop 1, the vortex has escaped out of the loop and is located on a pinning site. \textbf{c} After applying -14.6 mA through Loop 1 - the vortex was pulled back into the loop.  \textbf{d} An example of a minimal pull current measurement. The tip position is marked above a specific vortex (red circle). The current through the loop is set to each value and subsequently set to zero. The tip data (local field) is measured (black dots) until the the field value drops, a scan verifies that the vortex was indeed pulled into Loop 1. \textbf{e} A current map of the minimal current needed to pull vortices from identified pinning site into Loop 1, the minimal current value increases with distance from the loop center. \textbf{f} Same as d for Loop 2, the color scale is shared between e and f.}
    
\label{fig:Statistics}
\end{figure}

We now turn to the use of the Nb loops in elementary vortex manipulation. The application of current will exert an attractive or repulsive force between the loop and the vortex, depending on the current direction. We show here that we can pull a vortex into the center of a given loop, and push a vortex away from that loop.
To reduce vortex-vortex interaction, we initialize the system with low vortex density. In~Fig.~\ref{fig:Statistics}a we see the initial state where a vortex is centered within Loop 1. We then apply a current of I = 2.2 mA through Loop 1 (lower than $I_c \sim 15$ mA). In ~Fig.~\ref{fig:Statistics}b we image the state of the system following this operation. The vortex that was previously centered within Loop 1, was pushed away from the loop, and is now localized at a pinning site in the central area between the loops. We thus coin the application of a positive current as a \textit{push} operation. 

We demonstrate the opposite, \textit{pull} operation, by applying a negative current on the same loop. The vortex, which was pinned in the central area between the loops, is back to the center of Loop 1 (see Fig.~\ref{fig:Statistics}c). The pull operation requires a different current, depending on the pinning strength and the distance between the vortex and the loop. We determine the pull current using the following procedure: The current within a loop is set to a specific value, then set back to zero. The local field is then measured (black dots, Fig.~\ref{fig:Statistics}d) at the vortex location (red circle). At the minimal pull current, the local field at the vortex location drops. The images in the inset, taken before and after the vortex dislodged, confirm that the vortex has been pulled into Loop 1. 

While calibrating the push and pull operations, we typically acquire an image of the loop area after each current increment in the loops. For more sophisticated protocols (below) we station the SOT at the vortex location, and continuously monitor its signal until we find that the vortex has moved. We repeat this procedure to determine the minimal pulling current required to dislodge vortices residing at different locations. In Fig.~\ref{fig:Statistics}e,f we show a map of these minimal currents (color coded) needed to pull a vortex from each pinning site to Loops 1 and 2 respectively. These statistics demonstrate that each loop has a different operational pull range, likely a consequence of variations in the loop design. Important for what follows, is that each loop can pull vortices far enough to realize inter-loop vortex shuttling operations.

\begin{figure}[H]
\center

\includegraphics[width=\linewidth]{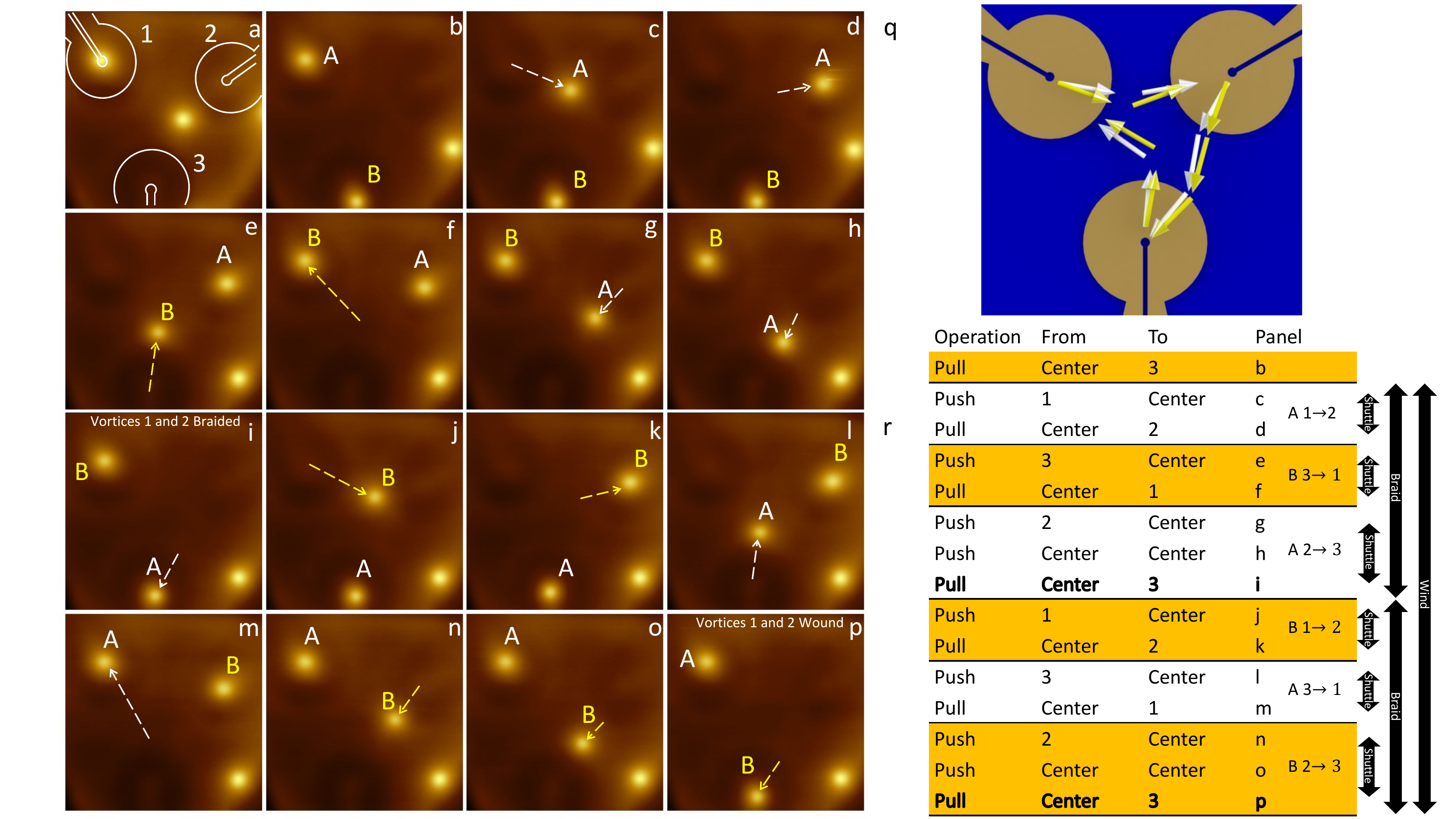}
\caption{\textbf{Vortex braiding - } \textbf{a} The initial state of the system after a field cool down using the loop magnetic field. \textbf{b} System initialized by pulling Vortex B into Loop 3. Loops 1 and 3 are populated, prepared for the vortex winding operation. Panels b-p are used to create the movie found in the following \href{https://youtube.com/shorts/O5biuFE1kDk}{link.} \textbf{c-p} Push and pull operations - circling vortices A and B around each other. \textbf{i} The two vortices have switched positions and have thus been braided. \textbf{p} Vortices have completed the full circle around each other, and have thus been wound. The detailed push and pull protocols are described in panel r. Each image is 6  $\times$ 6 $\mu$m$^2$, contains 128 pixels, and took $\sim$4 min to acquire. \textbf{q} A schematic of the vortices full route (A - white, B - yellow). \textbf{r} Table listing the operations performed in images \textbf{a-q}. Each line includes an operation (push / pull),  the initial and final locations (e. g. into Loop 3, from the central area) and the panel where the outcome is imaged. Horizontal lines highlight complete shuttle operations and bold text highlights a complete braiding operation. }
    
\label{fig:Braiding}
\end{figure}

We now discuss the combination of elementary push and pull operations to perform more complex protocols, including shuttle, braid and wind operations. A movie of the complete wind operation may be found in the following \href{https://www.youtube.com/shorts/O5biuFE1kDk}{link.} The initial state, consisting of two vortices in the field of view, created by field cooling is depicted in~Fig.~\ref{fig:Braiding}a. A subsequent pull operation with Loop 3 using $I=-2.3$ mA results in the configuration appearing in Fig.~\ref{fig:Braiding}b, where ``Vortex A'' resides in Loop 1, and ``Vortex B'' resides in Loop 3. 
From this initial state, we apply a sequence of operations in the order listed in Fig.~\ref{fig:Braiding}r. For each operation a current pulse of a few seconds was applied. We note that, in principle, this could have been fully automated and performed much faster as we discuss below. Subsequent to every operation, the state of the system is imaged in Fig.~\ref{fig:Braiding}b-p.
From now on, in order to track both vortices simultaneously we track their path using white (Vortex A) and yellow (Vortex B) dotted arrows. In Fig.~\ref{fig:Braiding}c, Vortex A is pushed from Loop 1 using 3.2 mA. In the picture, the vortex is located on a pinning site in the central area between the loops. In the table in Fig.~\ref{fig:Braiding}r, when a vortex resides in such a central pinning site it is designated as ``center''. 
Next, we pull Vortex A into Loop 2 using $I=-2.6$ mA in~Fig.~\ref{fig:Braiding}d. At this point, Vortex A has been guided from Loop 1 to Loop 2 using successive push and pull operations. We coin this a \textit{shuttle} operation. 

In the next few steps, Vortex B is shuttled from Loop 3 to Loop 1 (Fig.~\ref{fig:Braiding}e,f), and subsequently Vortex A is shuttled from Loop 2 to Loop 3 (Fig.~\ref{fig:Braiding}g-i). The state depicted in Fig.~\ref{fig:Braiding}i, reflects a full exchange of the two vortices - i.e. a braiding operation (noted by an arrow in Fig.~\ref{fig:Braiding}r). The next sequence of operations, ending in panel p completes a winding operation (noted by an arrow in Fig.~\ref{fig:Braiding}r). Thus, by a sequence of push-pull operations, we have demonstrated the capability of our device to carry out shuttle, braiding and winding operations - all of which are considered as fundamental in the toolbox of vortex-based quantum computation protocols~\cite{Nayak_RMP_2008,Ady2013}.

We now estimate the maximum operation frequency of such devices. The fundamental limiting factor is the viscous drag $\eta_0$ and the driving force that determines the vortex speed through $F_d=\eta_{0}v$. $\eta_0=\frac{\Phi_{0}}{2\pi\xi^2\rho_{n}} =5.55\cdot10^{-8}$ kg/m$\cdot$s is estimated from the Bardeen-Stephen model \cite{Bardeen1965} where $\Phi_0$ is the flux quantum, $\xi$ is the coherence length and $\rho_n$ is the normal state resistivity. This model proved to be a good approximation even at high speed~\cite{Embon2017}. The driving force is estimated from the force curve (see Supplementary Note 3). The vortex speed $v$ is calculated numerically along a straight line from the central area to the center of a loop. The average speed is 4.5 km/s. Such high velocity, comparable with the depairing velocity ($\sim 16$ km/s) is consistent with previous observations~\cite{Embon2017}. According to our calculation, the time of flight between two loops is of the order of 0.3 ns, limiting the operating frequency to about 3 GHz. This speed can be further increased if the device dimensions are shrunk.

For a fully integrated device, the vortex-control loops should operate in tandem with local vortex detection - done either using tunneling~\cite{Dvir2018}, or using RF local vortex detection. The latter could be realized using a local SQUID~\cite{Levenson-Falk2016}, fabricated above the loop center, or by using the vortex control loop as part of a pick-up loop which serves as part of a SQUID. As we show in Supplementary Note 4, the loop itself is sensitive to the presence of a vortex - as see in the $V(I)$ characteristics of the loop by a small variation in $I_c$.

Our results raise the question of coherent vortex motion, and whether vortices are moved adiabatically. Taking the velocity calculated above, we propose that the condition for adiabatic motion is keeping $hv/\xi$ smaller than a characteristic excitation energy scale. Which energy scale should be taken here, though, is not clear - whether it is the superconducting gap $\Delta$ ($\approx$ 1.26 meV for NbSe$_2$~\cite{Dvir2017}) or the gap between states bound to the vortex core.
In our device $hv/\xi \approx$ 3 meV at maximal speed - at the order of magnitude of $\Delta$, but far greater than the gap between Caroli-de-Gennes Matricon (CdGM) states in NbSe$_2$. 
Such protocols may hence be tested on materials with a sparse CdGM spectrum, such as \FTS~\cite{Machida2019}. 

It is also interesting to consider magnetic skyrmions, which have recently attracted attention within the spintronics community due to their potential applications \cite{Wiesendanger2016,Fert2017,Klaui2015}. 
Our technique could be used to manipulate these magnetic textures as has been suggested in~\cite{Cho2020} and done with MFM~\cite{Casiraghi2019}, due to the fact that magnetic force applied by the loops should behave the same. The exclusion zone mentioned in this paper should also be present with skyrmions and with any magnetic entity. Finally, our method should be compatible with hybrid structure that involves skyrmion-vortex pairs \cite{Nothhelfer2021}.

\section*{Methods}

\textit{Scanning SQUID-On-Tip Microscopy}
\newline
The SOT was fabricated using self aligned three step thermal deposition of Pb at cryogenic temperatures, as described in ref \cite{Zeldov2013}. The quartz tube was pulled to create a tip diameter of roughly $\sim$ 250 nm and the carefully adjusted deposition thicknesses resulted in a SQUID with a critical current of 165 $\mathrm{\mu}$A at zero field. The relatively large diameter tip allows for high magnetic field sensitivity and a slight asymmetry in the Josephson junctions shifts the interference pattern of the SQUID, resulting in finite magnetic field sensitivity at 0 applied field\cite{Zeldov2013}. This is crucial in order to conduct the experiment at low enough field to avoid overcrowding the sample with vortices.

\textit{Loop Fabrication}
\newline
100 nm Thick Nb was sputtered on a Si/SiO$_2$ chip. The Loops were written using standard E-beam lithography techniques on 950-A5 PMMA. After development, the Nb was etched using reactive ion etching (RIE). After the etching process the Nb layer was coated with 15 nm \alum\ by ALD at which point the NbSe$_2$ flake was transferred on top.

\section*{Acknowledgements}

We thank Ady Stern, Yuval Oreg, Eli Zeldov, Maxim Khodas, Nadav Katz, Shahal Ilani, Valla Fatemi and Tom Dvir for insightful discussions. HS acknowledges funding by Israel Science Foundation Quantum Initiative grant 994/19, Israeli Science Foundation grant 861/19. YA acknowledeges funding by the European Research Council grant No. 802952.
\section*{Contributions}

I.K., H.S. and Y.A. conceived the experiment. 
I.K. fabricated the vortex manipulation device. A.G. and N.F. fabricated the SOT. A.N. designed the SOT microscope. A.D.B. sputtered Nb for device fabrication. Measurements were performed by I.K. and Y.A., the paper was written by I.K., H.S. and Y.A. with input from all the co-authors.

\section*{Supplementary Information}

\subsection*{Supplementary Note 1. Device Details}

We measure the flake thickness to be 50 nm using an AFM. The scan was taken after the measurements. The measurements ended when the tip crashed on the sample surface, its debris is visible in the image. The flake is flat in the region of interest of the experiment.

\subsection*{Supplementary Note 2. Design Considerations}

To reach an effective localization and control of vortices, required for sophisticated shuttling and winding protocols, we had to take into account several design considerations. First, the inner diameter should be comparable to $\lambda$ in order to prevent multiple vortices from entering the loop but larger than $\xi$ to accommodate a vortex inside. The outer diameter should be large to enlarge exclusion zone but not too large on the scale of $\lambda$ to avoid vortices entering the loop. We found that 200 nm  for the inner diameter and 2.6 $\mu$m for the outer diameter are satisfactory parameters. The distance between different loops depends on the distance over which the force decays. This is set by the width of the loop ($r_{out}-r_{in}$). 
The control layer should be thicker than the target layer in order to increase the contrast in energy cost of locating a vortex in the exclusion zone. The loop thickness should not be much thicker than the target layer, to keep the target device planar. The control layer material should be a superconductor with  $\lambda$ as small as possible to enhance the exclusion zone effect. Smaller $\lambda$ or larger superfluid density has the added benefit of enhancing the critical current of the loops which will enhance the maximal force applied on the vortices.
The loops have to break symmetry in one axis, as shown in Fig. 1b, so the net force will point along the line of symmetry justifying the fact that all the loop leads points toward the central area in Fig.~1a and c. Finally, the target layer should be a material with relatively low pinning forces to allow the vortex manipulation.

\subsection*{Supplementary Note 3. Calculation of The Force Exerted By The Loop}

We calculate the force applied on the vortex by the loop, by numerically solving the integral $\textbf{F}=\int(\textbf{j} \times \textbf{B}) d^{3}x$. We assume a homogeneous current density $\textbf{j}$ within the loop while accounting for the loop shape. The magnetic field is given by the following expression $B_{z}=-\frac{\Phi_{0}}{2\pi\lambda^2} K_{0}(\frac{\sqrt{(x-d)^2+y^2}}{\lambda})$ - where d is the distance between the loop center and the vortex core. Here - $\Phi_{0}$ is the flux quantum, $\lambda$ the penetration length and $K_0$ is the MacDonald function. We plot the results for the largest obtained supercurrent (14 mA) in Supplementary Fig.~\ref{fig:Force}a. We note that this model is only valid when the vortex is far from the loop, and the loops Meissner screening is small. The loop outer radius is marked in Supplementary Fig.~\ref{fig:Force}a as a dashed line. From this force we can deduce the vortex speed, through $F_d=\eta_{0}v$ (Supplementary Fig.~\ref{fig:Force}b). Where $\eta_0=\frac{\Phi_{0}}{2\pi\xi^2\rho_{n}}$ is the Bardeen-Stephen viscosity \cite{Bardeen1965}. Here we used $\xi=7$nm \cite{Nader2014} and $\rho_{n}=0.025$ m$\Omega\cdot$cm \cite{Bawden2016}. We calculate the position vs time curve displayed in Supplementary Fig.~\ref{fig:Force}c.

\subsection*{Supplementary Note 4. Loop as a Vortex Detector}

Here, we measure the loop's $I_c$ dependence on the presence of a vortex in the loop (occupation state). In~Figure \ref{fig:Ic}a we present two I-V curves (2-probe with a 100 Ohm resistor in series) of loop 1. The black curve represents the occupied state measurement while the red curve represents the unoccupied measurement, both measurements are performed with the same polarity - with voltage ramping up from zero. One can see a difference in critical current values ($I_c$) at which resistance of the loop becomes finite. We repeat this measurement many times and draw three histograms In~Figures\ref{fig:Ic}b-d. Surprisingly, we note that $I_c$ increases when the loop is occupied. We attribute this effect to the magnetic flux being focused at the center of the ring, reducing the effective field on the loop and therefore increasing $I_c$. It should be noted that this method of sensing is invasive since the large current also generates a force on the surrounding vortices. Also, this measurement requires crossing the critical current, introducing heat to the system.  Nevertheless, if one applies a pulling force and find the loop to be unoccupied, one can conclude that the loop was not occupied even prior the the readout.

\renewcommand{\figurename}{Supplementary Figure}
\setcounter{figure}{0} 

\begin{figure}[H]
\center

\includegraphics[]{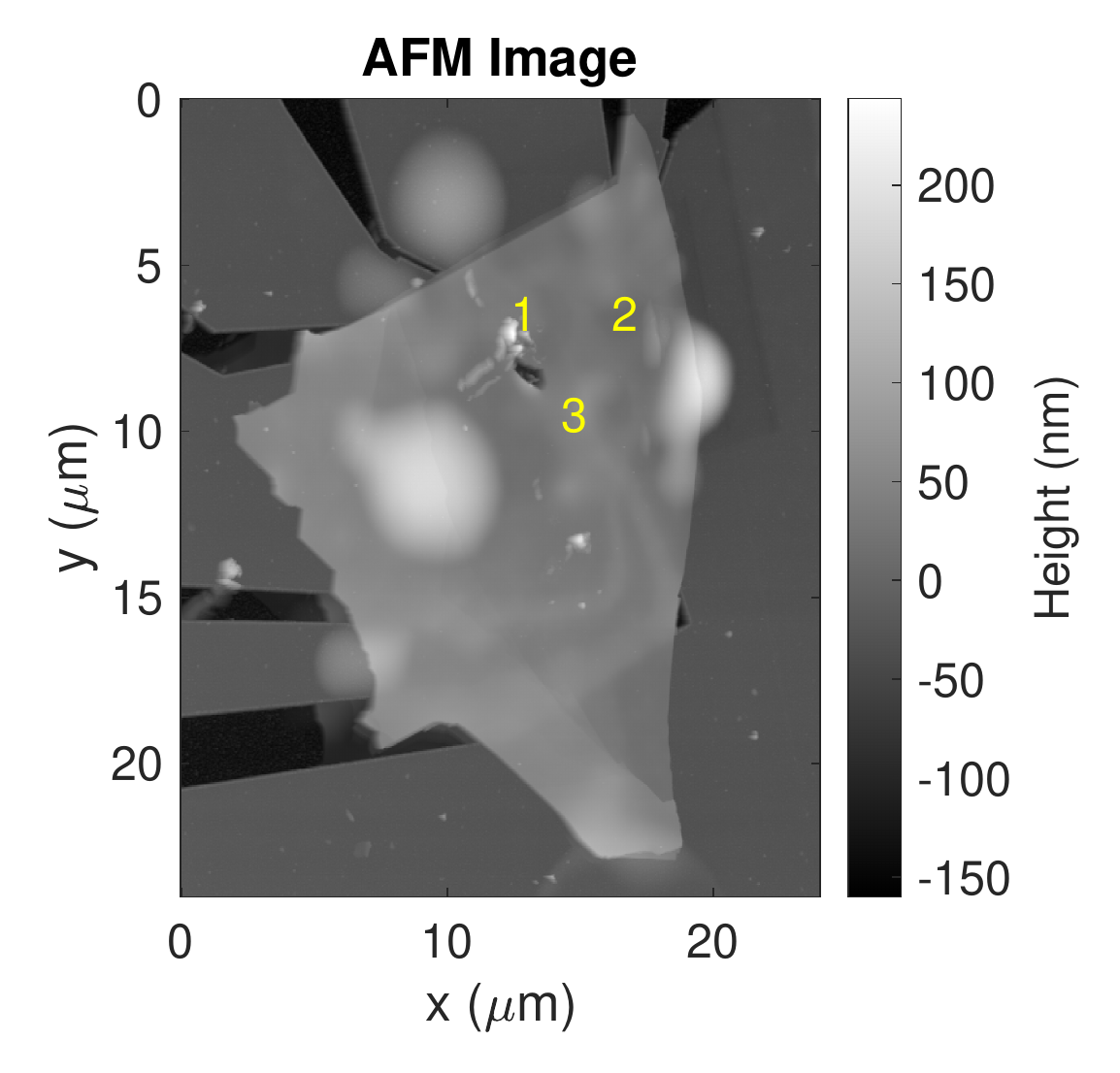}
\caption{AFM image of the NbSe$_2$ flake. The flake resides above the loop array. Debree from the crashed tip is visible to the left of loop 3. The flake is flat in the region of interest. }
    
\label{fig:AFM}
\end{figure}

\begin{figure}[H]
\center

\includegraphics[width=\linewidth]{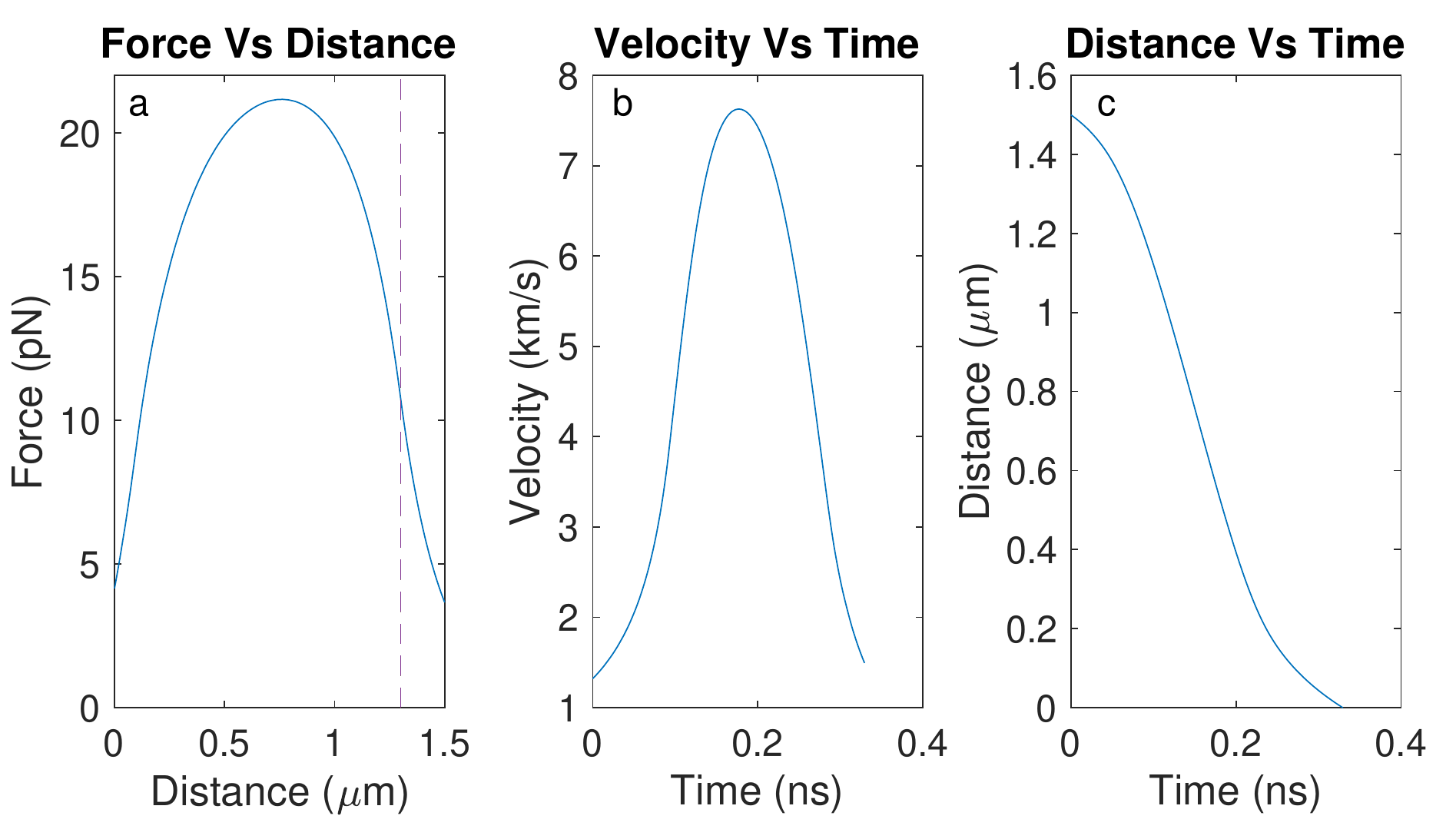}
\caption{Calculation of the vortex motion. \textbf{a}. Force vs Distance. This curve is simulated for a homogeneous current distribution within the loop, with the loop's symmetry broken along the axis connecting it's center to the vortex. \textbf{b}. Velocity vs time. We calculate the velocity vs time curve, by propagating the vortex through small time steps and quering the force vs distance curve for the velocity value at each position, through the Bardeen Stephen relation. \textbf{c}. same calculation as b, but the distance curve is displayed.}
\label{fig:Force}
\end{figure}

\begin{figure}[H]
\center

\includegraphics[width=\linewidth]{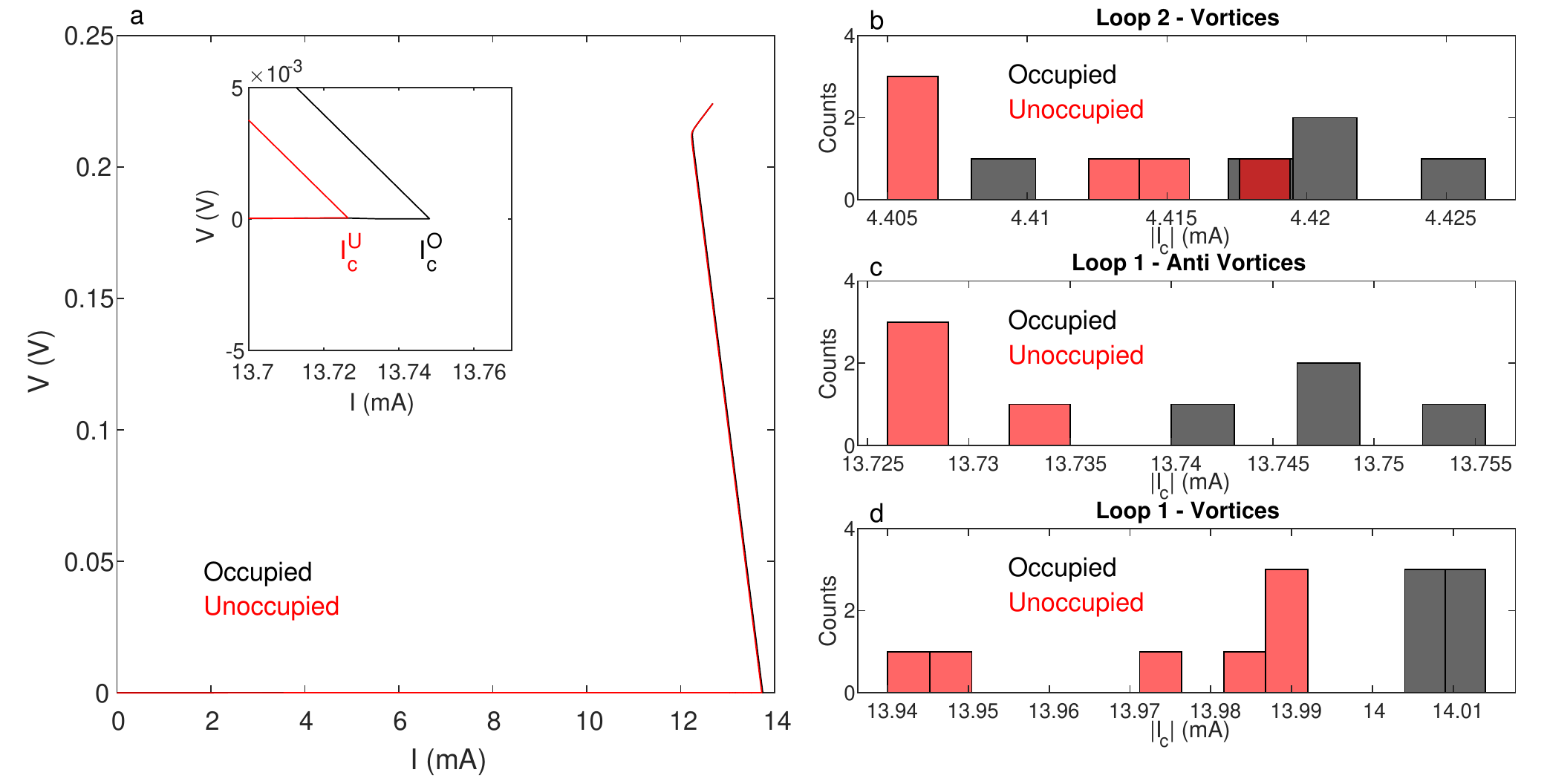}
\caption{\textbf{Critical current of the loop as a function of its occupation state} \textbf{a}. Typical $I_c$ measurement for an occupied (black curve) and unoccupied (red curve) loop 1. \textbf{b-d}. Histograms of measurements on loops 1 and 2, including vortices \textbf{b-c} and anti-vortices \textbf{d} representing two orientations of the external field. Notably, the critical current varies by ~20 $\mu$A depending on the occupation state of the loop.}
    
\label{fig:Ic}
\end{figure}

\providecommand{\latin}[1]{#1}
\makeatletter
\providecommand{\doi}
  {\begingroup\let\do\@makeother\dospecials
  \catcode`\{=1 \catcode`\}=2 \doi@aux}
\providecommand{\doi@aux}[1]{\endgroup\texttt{#1}}
\makeatother
\providecommand*\mcitethebibliography{\thebibliography}
\csname @ifundefined\endcsname{endmcitethebibliography}
  {\let\endmcitethebibliography\endthebibliography}{}

\end{document}